# THE CHARACTERISTIC NOISE INDUCED BY THE CONTINIOUS MEASUREMENTS IN CLASSICAL OPEN SYSTEMS

E.D. Vol
*B. Verkin Institute for Low Temperature Physics and Engineering of National Academy of Sciences of Ukraine,
Lenin ave., 47, Kharkov 61103, Ukraine*

Abstract

We proposed the modified version of quantum-mechanical theory of continuous measurements for the case of classical open systems. In our approach the influence of measurement on evolution of distribution function of an open system is described by the Fokker-Planck equation of a special form. The diffusion tensor of this equation is uniquely defined by a type of the measured quantity. On the basis of the approach proposed the stationary states of the linear dissipative systems, induced by measurements in them, are considered. Also we demonstrate on the simple example, how in the conservative system, consisting of noninteracting parts, measurement of the integral of motion results in relaxation to the quasi-thermodynamic equilibrium between parts of the system. The "temperature" of such state is determined by energy of the system and by the mean value of measured integral of motion.

It is well known that the role of measurement in quantum mechanics is much broader than in the classical physics, where it's role only passive and consists in obtaining by an experimenter the necessary information about the observed system. It is very essential to emphasize that in the last case all information about the system may be obtained without any disturbance of it's state by the measuring device(meter).Conversely in the quantum mechanics according by to the uncertainty principle it is impossible to eliminate back reaction of the meter on the state of measured system. Nevertheless using the quantum theory of continuous measurements (see the review [1] and references in it), it is possible not only to calculate the influence of measurement on a state of a system but also to use this influence for state monitoring. On the other hand because there is no impenetrable border between classical and quantum world, the natural question emerges: whether it is possible on the basis of classical concepts to evaluate the influence of a meter on a state of measured system whatever small this influence would be? In this paper we propose the simple approach to answer this question.

As one knows, the description of evolution of an open quantum system $Q$ in Markov approximation is given by Lindblad equation for density matrix $\hat{\rho}$ of this system [2]:

$$\frac{d\hat{\rho}}{dt} = -\frac{i}{\hbar}\left[\hat{H},\hat{\rho}\right] + \sum_i \left[\hat{R}_i\hat{\rho},\hat{R}_i^+\right] + \left[\hat{R}_i,\hat{\rho}\hat{R}_i^+\right], \qquad (1)$$

where $\hat{H} = \hat{H}^+$ and $\hat{R}_i, \hat{R}_i^+$ are set of operators describing both internal dynamics of system $Q$ and its connection with an environment.

On the other hand, the description of behavior of classical open system $C$ effected by additional stochastic forces (noise) can be obtained in Markov approximation in the framework of the Fokker-Planck equation for its distribution function $f(q,t)$ [3]:

$$\frac{\partial f}{\partial t} = -\frac{\partial}{\partial q_i}(K_i f) + \frac{\partial^2}{\partial q_i \partial q_k}(D_{ik} f), \qquad (2)$$

where: $q \equiv \{q_i\}$ - set of the coordinates describing a state of a system $C$, $K_i(q)$ - drift vector of $C$, determined by the equations of motion: $\frac{dq_i}{dt} = K_i(q)$, and $D_{ik}(q)$ - diffusion tensor which form is defined by the correlation tensor of stochastic forces field. Assume now that quantum system $Q$ has classical analog $C_Q$. It looks enough plausible that between these two descriptions should exist close connection. Such connection should allow one starting from a set of operators $\hat{H}, \hat{R}_i, \hat{R}_i^+$ for $Q$, in a limit $\hbar \to 0$, by the regular procedure to determine the expressions for $K_i(q)$ and $D_{ik}(q)$ of the classical system $C_Q$. Such correspondence between classical and quantum description for open Markov system was investigated in the paper of the author [4]. It has been shown, that in the first order on $\hbar$ Lindblad equation (1) for $\hat{\rho}$ turns to the classical Liouville equation for $f(q,t)$, i.e. actually to the equation (2) but without diffusive term. Meanwhile in [4] author did not take into account possible measurements produced under $Q$ (because their effect in the first order on $\hbar$ is strictly equal to zero). The main goal of the present paper is to point out that taking into account the influence of measurements on $C_Q$ evolution results in second order on $\hbar$ to the Fokker-Planck equation for $f(q,t)$ with diffusion tensor which form is uniquely determined by measured quantity. Let us demonstrate this statement at the example of the open system $Q_1$ with one degree of freedom which evolution is prescribed by measurement of physical quantity (observable) $O$. According to the quantum mechanics, the hermitian operator $\hat{O}$ corresponds to observable $O$. The equation for evolution of density matrix $\hat{\rho}$ of system $Q_1$ under the continuous measurement of $O$ is (see [1]):

$$\frac{d\hat{\rho}}{dt} = \frac{\gamma}{2}\left[\hat{O}\hat{\rho},\hat{O}\right] + \frac{\gamma}{2}\left[\hat{O},\hat{\rho}\hat{O}\right] = -\frac{\gamma}{2}\left[\hat{O},\left[\hat{O},\hat{\rho}\right]\right], \qquad (3)$$

where $\gamma$ is coupling constant between the meter and the measured system.

Further, following the method of [4], we make passage to the limit $\hbar \to 0$. Under such passage one must replace the density matrix $\hat{\rho}$ of $Q_1$ by it's classical counterpart - $f(q,p,t)$ of system $C_{Q_1}$, and commutators in the r.h.s. of (3) by Poisson brackets according to the Dirac rule: $[\hat{A},\hat{B}] \to i\hbar\{A,B\}$, where $\{A,B\} = \frac{\partial A}{\partial q}\frac{\partial B}{\partial p} - \frac{\partial A}{\partial p}\frac{\partial B}{\partial q}$ and $A(q,p)$ $B(q,p)$ are classical analogs of operators $\hat{A}$ and $\hat{B}$. Calculating double commutator according to this rule we easily come to the desired equation for distribution function $f(q,p,t)$ of $C_{Q_1}$:

$$\frac{\partial f}{\partial t} = \frac{\gamma\hbar^2}{2}\{O,\{O,f\}\} = \frac{\gamma\hbar^2}{2}\frac{\partial}{\partial x_i}(D_{ik}\frac{\partial f}{\partial x_k}), \tag{4}$$

where diffusion tensor $D_{ik}$ is determined by the measured quantity $O(q,p)$ ($O(q,p)$ is the classical counterpart of observable $\hat{O}$) with the help of relation:

$$D_{ik}(x) = \varepsilon_{il}\varepsilon_{km}\frac{\partial O}{\partial x_l}\frac{\partial O}{\partial x_m}. \tag{5}$$

In expressions (4) and (5) we use notation: $x_1 = q, x_2 = p$ and $\varepsilon_{ik} = \begin{pmatrix} 0 & 1 \\ -1 & 0 \end{pmatrix}$ is the antisymmetric tensor the second rank ($i,k = 1,2$).

The equation (4) can be easily reduced to the standard form of the Fokker-Planck equation [3]:

$$\frac{\partial f}{\partial t} = -\frac{\partial}{\partial x_i}(B_i f) + \frac{\gamma\hbar^2}{2}\frac{\partial^2}{\partial x_i \partial x_k}(D_{ik} f), \tag{6}$$

where $B_i \equiv \frac{\gamma\hbar^2}{2}\frac{\partial D_{ik}}{\partial x_k}$ is the drift of system $C_{Q_1}$ caused by influence of measurement.

In view of importance of eq. (4) and eq. (5) further we bring additional argument in behalf of these equations adequately describe the measurement process influence on distribution function of classical system. With this purpose let us introduce the quantity $\tilde{O}(q,p)$ such that variables $O(q,p)$ and $\tilde{O}(q,p)$ are complemented each other. It means that their Poisson bracket is equal to one. The equation (4) written in variables $O$ and $\tilde{O}$ can be represented as:

$$\frac{\partial f}{\partial t} = \frac{\partial}{\partial x_i}\left(D_{ik}\frac{\partial f}{\partial x_k}\right) \equiv \frac{\partial^2}{\partial \tilde{O}^2}f(O,\tilde{O},t). \tag{7}$$

(To simplify the notation we use for a moment system of units with $\frac{\gamma\hbar^2}{2} = 1$).

It is implied from eq. (7) that the influence of measurement $O$ on distribution function $f(O,\tilde{O},t)$ expressed in variables $O$ and $\tilde{O}$ leads to diffusion of $f(O,\tilde{O},t)$ only on variable $\tilde{O}$. The value of distribution function for $f(O,\tilde{O},t)$ in arbitrary time is given by well-known expression:

$$f(O,\tilde{O},t) = \frac{1}{(2\pi t)^{1/2}} \int d\tilde{O}_1 f(O,\tilde{O}_1,t=0) \exp\left(-\frac{(\tilde{O}-\tilde{O}_1)^2}{4t}\right). \tag{8}$$

It follows from the expression (8) that $\lim_{t\to\infty} f(O,\tilde{O},t) \equiv f_\infty(O,\tilde{O})$ is the function depending only on $O$. Therefore $f_\infty(O)$ may be regarded as distribution function for values of observable $O$ received as a result of continuous measurement of $O$. Such interpretation completely consistent with quantum theory (see [1]) in which measurement of $\hat{O}$ results in exponential decay of non-diagonal on $O$ matrix elements of $\hat{\rho}(t)$.

Let us turn now to the study of concrete examples showing influence of noise, induced by the measurements, on behavior of classical system.

As the first example we consider the evolution of linear open system $\mathcal{L}$ under the noise induced by continuous measurements in it. For simplicity we suppose that system $\mathcal{L}$ has only one degree of freedom, and coordinates $x_1$ and $x_2$ describing it's state are dimensionless. The equations of motion for variables $x_1$ and $x_2$ of such system can be written as:

$$\frac{dx_i}{dt} = A_{ik} x_k. \tag{9}$$

We assume that elements of constant matrix $\hat{A} \equiv \begin{pmatrix} a & b \\ c & d \end{pmatrix}$ are independent on $x_1$, $x_2$ and time, and satisfy two additional restrictions: 1) $\operatorname{Tr} A \equiv t_{\hat{A}} \equiv a+d < 0$ and 2) $\operatorname{Det}\hat{A} \equiv d_{\hat{A}} \equiv ad-bc > 0$. These restrictions provide exponentional decay of solutions (9) when $t \to \infty$. As established above the evolution of distribution function $f(x_1,x_2,t)$ of $\mathcal{L}$ when both drift (9) and continuous measurement are took into account described by the Fokker-Planck equation:

$$\frac{df}{dt} = -\frac{\partial}{\partial x_i}(A_{ik} x_k f) + D_{ik} \frac{\partial^2 f}{\partial x_i \partial x_k}. \tag{10}$$

We assume that for linear system the tensor of diffusion $\hat{D} = \begin{pmatrix} D_1 & D \\ D & D_2 \end{pmatrix}$ corresponding to the measurement in $\mathcal{L}$ by virtue of (5) satisfies to the condition:

$$d_{\hat{D}} = D_1 D_2 - D^2 = 0.\tag{10a}$$

Now we begin to study the stationary states of $\mathcal{L}$ induced in it by process of measurement. The method used for this purpose as a matter of fact is the same which used in the statistical physics under considering the fluctuations of physical quantities near equilibrium state (see [5]). We are looking for stationary solutions of the Fokker-Planck equation (10) in a standard form: $f(x_1, x_2) \sim \exp(S)$, where $S(x_1, x_2) = -\frac{1}{2}\beta_{ik} x_i x_k$ is a negative definite quadratic form of $x_1$ and $x_2$ which plays a role of entropy for stationary state. Substituting this expression for $f(x_1, x_2)$ into eq. (10), and, equating the coefficients at identical powers of variables $x_1$ and $x_2$, we obtain two equations for unknown symmetric matrix $\hat{\beta} \equiv \beta_{ik}$:

$$\operatorname{Tr} \hat{A} = -\operatorname{Tr} \hat{D}\hat{\beta},\tag{11a}$$

$$-\frac{\hat{\beta}\hat{A} + \hat{A}^t \hat{\beta}}{2} = \hat{\beta}\hat{D}\hat{\beta},\tag{11b}$$

where $\hat{A}^t$ is a matrix transposed to $\hat{A}$. It is easy to see that (11a) and (11b) are equivalent to the single equation for a matrix $\hat{\beta}^{-1}$, reciprocal to $\hat{\beta}$:

$$\hat{A}\hat{\beta}^{-1} + \hat{\beta}^{-1}\hat{A}^t = -2\hat{D}.\tag{12}$$

Note, that the matrix equation (12) allows one to obtain solution for $\hat{\beta}^{-1}$ in the case when matrix $\hat{A}$ has arbitrary dimension $N \times N$ (see [6]). However, here we are interested in only the situation when $N = 2$. In this case the expression for $\hat{\beta}^{-1}$ can be presented as:

$$\hat{\beta}^{-1} = \frac{(t_{\hat{A}}^2 + d_{\hat{A}})}{t_{\hat{A}} d_{\hat{A}}} \hat{D} + \frac{1}{d_{\hat{A}}}(\hat{A}\hat{D} + \hat{D}\hat{A}^t) - \frac{1}{t_{\hat{A}} d_{\hat{A}}} \hat{A}\hat{D}\hat{A}^t.\tag{13}$$

(We remind that $t_{\hat{A}} \equiv \operatorname{Tr} \hat{A}$ and $d_{\hat{A}} \equiv \operatorname{Det} \hat{A}$). It is known from the fluctuation theory (see [5]), that elements of a matrix $\hat{\beta}^{-1}$ coincide with the second moments of coordinates $x_1$ and $x_2$ in a stationary state. Therefore using (13) one can write down explicit expressions for these moments with the help of known elements of matrixes $\hat{A}$ and $\hat{D}$:

$$\overline{x_1^2} = \hat{\beta}_{11}^{-1} = \frac{(bc - ad - d^2)D_1 + 2bdD - b^2 D_2}{(a+d)(ad - bc)},\tag{14a}$$

$$\overline{x_1 x_2} = \hat{\beta}_{12}^{-1} = \frac{cdD_1 - 2adD + abD_2}{(a+d)(ad-bc)}, \tag{14b}$$

$$\overline{x_2^2} = \hat{\beta}_{22}^{-1} = \frac{-c^2 D_1 + 2acD + (bc - a^2 - ad)D_2}{(a+d)(ad-bc)}. \tag{14c}$$

Now having in hands expression for $\hat{\beta}_{ik}^{-1}$ and hence for the entropy $S(x_1, x_2)$ we can to construct "thermodynamics" of measurement process for linear open systems. By analogy to usual thermodynamics we define thermodynamic forces $X_i$ as:

$$X_i = \frac{\partial S}{\partial x_i} = -\beta_{ik} x_k. \tag{15a}$$

Coordinates $x_1$ and $x_2$ describing system state are expressed by forces $X_i$ as:

$$x_i = -\beta_{ik}^{-1} X_k. \tag{15b}$$

It is convenient to introduce the kinetic matrix $\hat{L} = -\hat{A}\hat{\beta}^{-1}$ and by means of it to write down the equations of motion for $\mathcal{L}$ (9) in the standard form as the connection between "flows" $j_i \equiv \frac{dx_i}{dt}$ and "forces" $X_i$:

$$j_i = A_{ik} x_k = -(\hat{A}\hat{\beta}^{-1})_{ik} X_k = L_{ik} X_k. \tag{16}$$

Comparing expression of $\hat{L}$ with the eq. (12), we obtain the relation:

$$\hat{L} + \hat{L}^+ = 2\hat{D}. \tag{17}$$

It is well-known that in linear nonequilibrium thermodynamics the kinetic matrix $\hat{L}$ is symmetric (if magnetic field is absent), i.e. $\hat{L} = \hat{L}^t$. This important result for the first time obtained by Onsager [7] follows from the symmetry of equations of motion with respect to time inversion. In the case of arbitrary open linear system $\mathcal{L}$ the equations of motion (9) obviously do not possess such symmetry. It is interesting to note that under definite restrictions on a measured quantity (which are determined by the drift matrix $\hat{A}$) kinetic matrix $\hat{L}$ turns out to be symmetric. Let us find these conditions in explicit form. For this purpose we substitute expression for $\hat{\beta}^{-1}$ from eq. (13) into definition of kinetic matrix $\hat{L} = -\hat{A}\beta^{-1}$ and after simple algebra obtain the following relation:

$$\hat{L} = \hat{D} + \frac{1}{t_A}(\hat{A}\hat{D} - \hat{D}\hat{A}^t). \tag{18}$$

From eq. (18) follows that the matrix $\hat{L}$ becomes symmetric under condition: $\hat{A}\hat{D} = \hat{D}\hat{A}^t$ or when elements of matrixes $\hat{A}$ and $\hat{D}$ are connected as:

$$bD_2 - cD_1 + (a-d)D = 0. \tag{19}$$

Recollecting now there is general restriction (10a) on the elements $\hat{D}$ which corresponds to the measurement process we come as a result to the following conclusion. For arbitrary open linear system there is an observable the continuous measurement of which induces the stationary state of a system with symmetric matrix $\hat{L}$. It is interesting to note that the same measurement results in the maximal correlation between coordinates. Let us prove this statement. We introduce the coefficient of correlation between $x_1$ and $x_2$ by means of usual definition:

$$\eta = \frac{\overline{x_1 x_2}}{\left(\overline{x_1^2}\, \overline{x_2^2}\right)^{1/2}}. \tag{20}$$

(It is implied in (20) that $\overline{x_1} = \overline{x_2} = 0$). Using the known expressions for second moments (14a-14c) it is easy to obtain the relation:

$$\frac{1}{\eta^2} = \frac{\overline{x_1^2}\, \overline{x_2^2}}{\left(\overline{x_1 x_2}\right)^2} = 1 + (ad - bc)\frac{[bD_2 - cD_1 + (a-d)D]^2}{(cdD_1 - 2adD + abD_2)^2}. \tag{21}$$

Comparing (21) with condition (19) we come to the declared result: symmetry of kinetic matrix $\hat{L}$ leads to equality $|\eta| = 1$, i.e. to the maximal correlation between $x_1$ and $x_2$ and vice versa. The sense of the result obtained becomes almost evident if we pass to variables $O$ and $\tilde{O}$ ($O$ -the measured value and $\{O, \tilde{O}\} = 1$). In these variables, directly connected with measurement, tensor of diffusion has the simple form: $\hat{D} = \begin{pmatrix} 0 & 0 \\ 0 & 1 \end{pmatrix}$ (see (7)) and condition of symmetry for matrix $\hat{L}$ looks as $b = 0$. The equations of motion in variables $O$ and $\tilde{O}$ have the form:

$$\frac{dO}{dt} = aO, \quad \frac{d\tilde{O}}{dt} = cO + d\tilde{O}. \tag{22}$$

The corresponding Fokker-Planck equation for distribution function $f(O, \tilde{O})$ of stationary state may be written as:

$$\frac{\partial}{\partial O}(aOf) + \frac{\partial}{\partial \tilde{O}}\left[(cO + d\tilde{O})f\right] = \frac{\partial^2 f}{\partial \tilde{O}^2}. \tag{23}$$

As one can easy to see the eq. (23) has normalized solution of the form:

$$f(O,\tilde{O}) = \sqrt{\frac{|d|}{2\pi}} \delta(O) \exp\left(-\frac{|d|\tilde{O}^2}{2}\right). \qquad (24)$$

Thus, under the condition $\hat{A}\hat{D} = \hat{D}\hat{A}^t$, $f(O,\tilde{O})$ turns out to be proportional to delta-function of measured quantity. It means the "freezing" of the observed quantity, which automatically results in the maximal correlation between coordinates of $\mathcal{L}$ and to the symmetry of kinetic matrix $\hat{L}$. One can say that under condition (19) we have analog of quantum Zeno effect (see [8]) in classical open system.

Let us discuss briefly the possibility of experimental observation the effects connected with influence of measurement on behaviour of macro- (meso-) scopical system. Two main obstacles can hinder such observation: 1) smallness of measurement noise proportional as we saw, to $\hbar^2$ and 2) unavoidable presence at experiment of extraneous noise of the different nature (thermal, shot and so on), which can suppress effects connected with measuring noise. The first obstacle is essential mainly for linear systems. Really, it is well-known, that in nonlinear system not far from bifurcation point even weak external noise can result in qualitative changing of the system state (see e.g. [9]). The simple example of such bifurcation under influence of measuring noise is considered in the appendix. The second obstacle, i.e. presence of some extraneous noise in the system, is more serious. We postpone the detailed analysis of this problem for future publications and note only that specific character of measuring noise and its selective influence on various physical quantities allows one to have a hope to select it from irrelevant noise of other nature.

In the final part of the paper we consider the interesting problem connected with influence of continuous measurement in composite system on behavior of its parts. To point out the basic physical idea and conclusions following from it without complicating our account with technical details we are restricted to considering the simplest example of such situation. Let us study the system $C$ consisting of two identical noninteracting harmonic oscillators with Hamiltonian:

$$H = H_1 + H_2 = \frac{p_1^2}{2m} + \frac{kx_1^2}{2} + \frac{p_2^2}{2m} + \frac{kx_2^2}{2}. \qquad (25)$$

The projection of angular momentum $M_z = x_1 p_2 - x_2 p_1$ is integral of motion because $\{M_z, H\} = 0$. Let us assume that continuous measurement of integral of motion $M_z$ occurs in this system. We are interested in how such measurement will affect behavior of oscillators 1 and 2. According to the approach proposed in the first part of the paper the evolution of distribution function of composite system $C$ may be described by the following Fokker-Planck equation:

$$\frac{\partial f}{\partial t} = -\frac{\partial}{\partial x_1}\left(\frac{\partial H}{\partial p_1}f\right) - \frac{\partial}{\partial x_2}\left(\frac{\partial H}{\partial p_2}f\right) + \frac{\partial}{\partial p_1}\left(\frac{\partial H}{\partial x_1}f\right) + \frac{\partial}{\partial p_2}\left(\frac{\partial H}{\partial x_2}f\right) + \kappa\{M_z,\{M_z,f\}\}, \qquad (26)$$

where $f(\Gamma_1,\Gamma_2,t) \equiv f(x_1,p_1;x_2,p_2;t)$ is distribution function of composite system $C$ and $\kappa = \frac{\gamma\hbar^2}{2}$ is a coupling constant of the meter with measured system $C$.

With the help of eq. (26) and using integration by parts one can obtain the dependence of mean value $\bar{A}(t)$ for any physical quantity $A(\Gamma_1,\Gamma_2,t)$ depending on time: $\bar{A}(t) = \int d\Gamma_1 d\Gamma_2 A(\Gamma_1,\Gamma_2,t) f(\Gamma_1,\Gamma_2,t)$ under the measurement of $M_z$. This dependence is given by following expression:

$$\frac{dA}{dt} = \overline{\{A,H\}} + \kappa \overline{\left(p_2\frac{\partial}{\partial p_1} - p_1\frac{\partial}{\partial p_2} + x_2\frac{\partial}{\partial x_1} - x_1\frac{\partial}{\partial x_2}\right)^2 A}. \qquad (27)$$

Using equality (27) one can write down the equations for all second moments, i.e. for values $\overline{x_i x_k}$, $\overline{p_i p_k}$ and $\overline{x_i p_k}$ ($i=1,2$) and for their linear combinations. Let us write for example the equation of motion for average energy $E_1 = \overline{\frac{p_1^2}{2m}} + \overline{\frac{kx_1^2}{2}}$ of oscillator 1. Using equality (27) we obtain:

$$\frac{dE_1}{dt} = 2\kappa(E_2 - E_1). \qquad (28a)$$

The similar equation of motion is also correct for oscillator 2:

$$\frac{dE_2}{dt} = 2\kappa(E_1 - E_2). \qquad (28b)$$

From eq. (28a) and (28b) expected result follows: energy of the composite system : $2E = E_1 + E_2$ under the measurement of $M_z$ is conserved. Moreover one can see that equalization of subsystems energies (thermalization) takes place. We want to point out that this thermalization connected exclusively with measurement of $M_z$ because dynamical interaction between oscillators 1 and 2 is strictly equal to zero.

Now let us write the equations of motion for the moments $\overline{x_1 p_2}$ and $\overline{x_2 p_1}$. Using eq. (27) we find:

$$\frac{d\overline{x_1 p_2}}{dt} = \frac{\overline{p_1 p_2}}{m} - k\overline{x_1 x_2} - 2\kappa(\overline{x_1 p_2} + \overline{x_2 p_1}), \qquad (29a)$$

$$\frac{d\overline{x_2 p_1}}{dt} = \frac{\overline{p_1 p_2}}{m} - k\overline{x_1 x_2} - 2\kappa(\overline{x_1 p_2 + x_2 p_1}). \tag{29b}$$

It follows from (29a) and (29b) that mean value of $M_z$: $\overline{M_z} \equiv \overline{x_1 p_2 - x_2 p_1} \equiv M$ does not depend on time and combined with total energy may be used for the characteristic of stationary state of the system during measurement $M_z$. The other equations of motion for second moments can be similarly obtained and the values of these moments may be determined in stationary state. Omitting trivial calculations, we present only the final results:

$$\frac{\overline{p_1^2}}{2m} = \frac{\overline{p_2^2}}{2m} = \frac{k\overline{x_1^2}}{2} = \frac{k\overline{x_2^2}}{2} = \frac{E}{2}, \tag{30a}$$

$$\overline{x_1 p_2} = \frac{M}{2}; \quad \overline{x_2 p_1} = -\frac{M}{2} \tag{30b}$$

$$\overline{x_1 x_2} = \overline{p_1 p_2} = \overline{x_1 p_1} = \overline{x_2 p_2} = 0. \tag{30c}$$

The knowledge of all second moments allows one to write down distribution function $f_C(\Gamma_1, \Gamma_2)$ for the stationary state of composite system $C$ in the form of Gauss distribution, thus that moments $\overline{x_i x_k}$, $\overline{p_i p_k}$ and $\overline{x_i p_k}$, determined by it coincide with known (30a), (30б), (30в). Let us represent $f_C(\Gamma_1, \Gamma_2)$ in a standard form $f \sim \exp(S)$, where $S(\Gamma_1, \Gamma_2) = -\frac{1}{2}\beta_{\alpha\beta} y_\alpha y_\beta$ is entropy of a stationary state of system $C$. We use following ordering of variables $y_\alpha$ ($\alpha = 1, 2, 3, 4$): $y_1 = x_1$, $y_2 = p_1$, $y_3 = x_2$, $y_4 = p_2$. The matrix $\hat{\beta}^{-1}$ reciprocal to matrix $\hat{\beta}$ which elements coincide with known moments can be represented as:

$$\hat{\beta}^{-1} = \begin{pmatrix} E/k & 0 & 0 & M/2 \\ 0 & mE & -M/2 & 0 \\ 0 & -M/2 & E/k & 0 \\ M/2 & 0 & 0 & mE \end{pmatrix}. \tag{31a}$$

In accordance with (31a) matrix $\hat{\beta}$ is equal to:

$$\hat{\beta} = \begin{pmatrix} mE & 0 & 0 & -M/2 \\ 0 & E/k & M/2 & 0 \\ 0 & M/2 & mE & 0 \\ -M/2 & 0 & 0 & E/k \end{pmatrix} \frac{1}{\dfrac{mE^2}{k} - \dfrac{M^2}{4}}. \tag{31b}$$

Now with the help (31b) one can write down the distribution function of composite system $f_C(\Gamma_1, \Gamma_2)$ in desired form:

$$f_C \sim \exp(-\beta(H - M_z \Omega)),  \qquad (32)$$

where notations: $\beta \equiv \dfrac{E}{\omega_0^2}\left(\dfrac{mE^2}{k} - \dfrac{M^2}{4}\right)^{-1}$, $\Omega \equiv \dfrac{M\omega_0^2}{2E}$ and $\omega_0 = \sqrt{k/m}$ are used.

Representation (32) for distribution function $f_C(\Gamma_1, \Gamma_2)$ is the basic result of this part of the paper. The small comments are necessary to it. First of all note that value of quantity $\dfrac{mE^2}{k} - \dfrac{M^2}{4}$ is more then zero, that is why parameter $\beta > 0$. This statement follows from inequality $\overline{(x_1^2)}\,\overline{(p_2^2)} \geq \overline{(x_1 p_2)}^2$ taking into account that $\dfrac{k\overline{x_1^2}}{2} = \dfrac{\overline{p_2^2}}{2m} = \dfrac{E}{2}$, and $\overline{x_1 p_2} = M/2$. The second remark is more essential. As one can see directly from (32) distribution function $f_C(\Gamma_1, \Gamma_2)$ of composite system may be written down in the form of Gibbs distribution with effective Hamiltonian $H_{eff} = H - \Omega M_z$. The effective temperature of such distribution $\kappa T_{eff} = E - \dfrac{M^2 \omega_0^2}{4E}$ ($\kappa$ is the Boltzmann constant) is determined by the total energy of the system and by mean value of the measured integral of motion.

It is worth to remind that both main effects: equalization of subsystems energies and setting of quasi-equilibrium Gibbs distribution occur in the system of noninteracting oscillators only due to process of measurement. The observation of this effects in macro- (meso-) scopical system would be the crucial argument in behalf of approach proposed in the present paper.

The author acknowledges L.A. Pastur for the discussion of the results of the paper and valuable comments.

Appendix

In this appendix we briefly consider simple example of bifurcation of a state of the macro- (meso-) scopical system, caused by measurements in it. Let us assume that nonlinear dynamical system is near to a bifurcation point, connected with emergence of auto-oscillations in it. According to [10] equation of motion for complex coordinate $z = x + iy$ (dimensionless quantities $x$ and $y$ describe a state of considered system in vicinity of a bifurcation point) can be written as:

$$\frac{dz}{dt} = z(i\omega + \varepsilon - c|z|^2), \qquad (A1)$$

where $i$ - complex unit and $\omega, c, \varepsilon$ are parameters describing the system, in particular, $\varepsilon$ defines distance up to a bifurcation point.

Let us introduce new variables: "action" - $j = \frac{x^2 + y^2}{2}$ and "angle" - $\varphi = -\arctan\frac{y}{x}$ and write down equations of motion for them:

$$\frac{dj}{dt} = 2\varepsilon j - 4cj^2; \qquad \frac{d\varphi}{dt} = \omega. \tag{A2}$$

The equations (A2) at $\varepsilon > 0, c > 0$ have the evident solution in the form of the limit cycle: $j = \frac{\varepsilon}{2c}, \varphi = \omega t + \varphi_0$. At the absence of any noise acting on the system, the Liouville equation for its distribution function $F_0(j,\varphi,t)$ has the stationary solution $F_0(j,\varphi) \sim \delta\left(j - \frac{\varepsilon}{2c}\right)$ describing movement of the system along its limit cycle.

Assume now that simultaneously with dynamics (A2) one produces the continuous measurement of the phase - $\varphi$. Accordingly to the approach proposed above the evolution of distribution function of the system $F(j,\varphi,t)$ satisfies to the following Fokker-Planck equation:

$$\frac{\partial F}{\partial t} = -\frac{\partial}{\partial j}(2\varepsilon j - 4cj^2)F - \omega\frac{\partial F}{\partial \varphi} + D\frac{\partial^2 F}{\partial j^2}, \tag{A3}$$

where $D = \frac{\gamma \hbar^2}{2}$. The stationary solution of (A3) is:

$$F(j,\varphi) \sim \exp\left(\frac{\varepsilon j^2}{D} - \frac{4cj^3}{3D}\right). \tag{A4}$$

We see that under continuous measurement of a phase $\varphi$ distribution function $F(j,\varphi)$ has two extrema: minimum at $j = 0$ and a maximum at $j = \frac{\varepsilon}{2c}$. The ratio of probabilities to detect system in these states is:

$$\frac{F_{max}}{F_{min}} = \exp\left(\frac{\varepsilon^3}{12Dc^2}\right). \tag{A5}$$

From relation (A5) follows that in the case when parameter $\frac{\varepsilon^3}{Dc^2} \leq 1$ there is noticeable possibility to observe the system out of its limit cycle. More precise evaluation of this possibility has meaning only under exact knowledge of parameters for concrete macro- (meso-) scopical system.